\documentclass[useAMS,usenatbib]{mn2e}
\usepackage{epsfig}

\title[Hibernation Revived by Weak Magnetic Braking]{Hibernation Revived by Weak Magnetic Braking}

\author[R. G. Martin and C. A. Tout]{Rebecca G. Martin and Christopher A. Tout \thanks{E-mail:
rgm32@cam.ac.uk; cat@ast.cam.ac.uk; }\\
University of Cambridge, Institute of Astronomy, The Observatories,
Madingley Road, Cambridge CB3 0HA\\}
\begin{document}

\date{}

\pagerange{\pageref{firstpage}--\pageref{lastpage}} \pubyear{2004}
\maketitle

\label{firstpage}

\begin{abstract}

Cataclysmic variables undergo periodic nova explosions during which a
finite mass of material is expelled on a short timescale. The system
widens and, as a result, the mass-transfer rate drops. This state of
hibernation may account for the variety of cataclysmic variable types
observed in systems of similar mass and period. In the light of recent
changes to the theory of nova ignition and magnetic braking we
investigate whether hibernation remains a viable mechanism for
creating cataclysmic variable diversity. We model the ratio of time
spent as dwarf novae (DNe) to nova-like systems (NLs). Above a
critical mass-transfer rate the system is NL and below it a DN. The
dominant loss of angular momentum is by magnetic braking but the rate
is uncertain. It is also uncertain what fraction of the mass accreted
is expelled during the novae. We compare the models of the ratios
against the period of the system for different magnetic
braking rates and different ejected masses with the ratio of the
number of observed NLs to DNe. We deduce that a rate of angular
momentum loss a factor of ten smaller than that traditionally assumed
is necessary if hibernation is to account for the observed ratios.
\end{abstract}


\section{Introduction}
Cataclysmic variables \citep{b6} are interacting close binary stars in
which a white dwarf is accreting from a companion. In the vast
majority of cases the companion is a low-mass main-sequence star or
red dwarf and the periods are measured in hours. The red dwarf is
transferring hydrogen-rich material to the white dwarf by Roche lobe
overflow at a rate, $10^{-11}$ to $10^{-8} M_\odot\,\rm yr^{-1}$. As
the hydrogen settles on the surface, the base of the layer becomes degenerate
and heats. When enough, $10^{-5}$ to $10^{-3} M_\odot$, has
accumulated it ignites. The degenerate conditions lead to a
thermonuclear runaway and a nova explosion. The system brightens by
ten or so magnitudes and most of the layer if not all of it together
with some of the underlying white dwarf material is blown off.

Though some of this material may fall back on to the white dwarf and
some of it may be accreted back by the red dwarf, a significant
fraction escapes carrying with it only the intrinsic angular momentum
of the white dwarf. The system expands such that $aM=\rm const$, where
$a$ is the separation and $M$ the total mass which has just
decreased. If the red dwarf had no atmosphere, that is a distinct
boundary just filling its Roche lobe, it would no longer do so and
mass transfer would cease. This is the idea of hibernation first
documented by \cite{b4}.  At that time they were motivated by the
apparent lack of cataclysmic variables in the solar neighbourhood when
compared with theoretical estimates and the frequency of classical
novae in M31.  The complete detachment during hibernation would render
the majority of cataclysmic variables unobservable at any given time.
\par
Because the red dwarf has a slightly extended atmosphere it actually
must overfill its Roche lobe in order to maintain an equilibrium
mass-transfer rate. The ratio of the atmosphere scale height to the
radius of the star is comparable to the ratio of the mass lost from
the system to its total mass, so the change in separation is not
sufficient to shut off mass transfer altogether but does reduce the
rate significantly. This change in rate could be responsible for the
diversity of cataclysmic variable types at a given period: dwarf novae
that undergo disc instability outbursts require a lower mean
mass-transfer rate than nova-likes that do not suffer the disc
instability
\citep{Faulk}. However the idea lost favour when it was realised that
the nova ejecta would temporarily form a common envelope around the
system and consequential angular momentum loss would actually cause
lower-period systems to shrink \citep{livio91}. Recent work by
\cite{Nele} indicated that the common-envelope phase is
very poorly understood and that in some cases systems with tenuous
envelopes can actually expand. Since we wish to compare the simplest
case we neglect any common-envelope phase in this work.
\par
A detailed analysis of the period distribution was made by
\cite{shafter92}.  He emphasized that the lack of dwarf novae between
3 and~$4\,$hr is difficult to reproduce theoretically with any
reasonable magnetic braking law.  He speculated that the solution
might lie in a better understanding of any of additional angular
momentum loss alongside mass transfer, weak but unnoticed magnetic
fields that disrupt the inner parts of the disc, correlation between
white dwarf mass and orbital period or hibernation.  Among these he
deduced that hibernation might prove the most promising and it is this
that we explore here in more detail.
\par
We investigate the hibernation scenario for cataclysmic variable
diversity in the light of two recent changes to the accepted model of
cataclysmic variables. These are
\begin{enumerate}
\item the relation between the mass of hydrogen that must be accreted
to ignite $M_{\rm ign}$ and the secular accretion rate \citep{TandB}
\item the possibility that magnetic braking, that drives the evolution
of cataclysmic variables, could be much weaker than previously thought \citep{And}.
\end{enumerate}
We consider hibernation in its simplest form to deduce what
combinations of $\dot J$, $M_{\rm ign}$ and $M_{\rm ej}$, the mass
actually ejected in the nova, could be consistent with the observed
variety of cataclysmic variable types.

\section{Simple Analytic Model}
We first consider a simple analytical model to get some idea of what
fraction of the time is spent in hibernation and then consider the
simplest numerical model that includes the pertinent effects of the
atmosphere and variations of $M_{\rm ej}$ and $\dot J$.

\subsection{Binary Systems}
The cataclysmic binaries we consider consist of a white dwarf, of mass
$M_1$, and a low-mass main-sequence star, of mass $M_2$ and radius
$R_2$.

The hydrostatic and thermal equilibrium radius of the main-sequence
star can be approximated by,
\begin{equation}
\frac{R_2}{R_{\odot}}=\frac{M_2}{M_{\odot}}.
\end{equation}

The Roche lobe radius $R_{{\rm L}i}$ is the radius of a sphere which encloses
the same volume as that enclosed by the last stable equipotential
surface around star~$i$.  The Roche lobe radius of star~2, for a restricted range
of mass ratio $q=M_2/M_1$, is
\begin{equation}
\frac{R_{\rm L2}}{a}=\frac{2}{3^{4/3}}\left(\frac{M_2}{M}\right)^{1/3},\qquad 0<q<0.8,
\end{equation}
where $M=M_1+M_2$ is the total mass of the system \citep{Pac}. The
simplicity of this formula makes it useful for analytic work within
the restricted but useful range of $q$. When $R_2>R_{\rm L2}$ mass is
transferred from the red dwarf to the white dwarf.

Kepler's third law, that relates the angular velocity $\Omega$ of each
star and the binary system to the total mass $M$ and the separation of
the two stars $a$, is
\begin{equation}
\Omega ^2 = \frac{GM}{a^3}.
\end{equation}
When the red dwarf is filling its Roche lobe, $R_2=R_{\rm L2}$, we combine
equations~(1),~(2) and~(3) to find
\begin{equation}
P= \frac{9 \pi}{(2G)^{1/2}}\left(\frac{R_{\odot}}{M_{\odot}}\right)^{3/2}  M_2,
\end{equation}
where $P=2 \pi / \Omega$ is the orbital period. So in this model the
orbital period depends only on the mass of the main-sequence star when
$R_2=R_{\rm L2}$.

We assume the spin angular momentum of the stars is negligible
compared to the orbital angular momentum. Hence the total angular
momentum of the system is
\begin{equation}
J=M_1a_1^2\Omega+M_2a_2^2\Omega,
\end{equation}
where $a_1$ and $a_2$ are the distances from star~1 and star~2
respectively to the centre of mass of the system. We can rearrange
\begin{equation}
a_1M_1=a_2M_2 \quad \rmn{and} \quad a=a_1+a_2
\end{equation}
to find
\begin{equation}
a_1=\frac{M_2}{M}a,
\end{equation}
and similarly for $a_2$. Substituting into equation~(5) we have
\begin{equation}
J=\frac{M_1M_2}{M}a^2\Omega .
\end{equation}
Using Kepler's law, equation~(3), and equation~(8) we can rearrange to get the separation 
\begin{equation}
a=\frac{J^2M}{GM_1^2M_2^2}
\end{equation}
and the angular velocity
\begin{equation}
\Omega = \frac{G^2M_1^3M_2^3}{J^3M}
\end{equation}
in terms of the three independent variables, $J$, $M_1$ and $M_2$.

\subsection{Nova Eruption}

When a critical mass $\Delta m$, which we examine later, has
accumulated on the surface of the white dwarf star, thermonuclear
reactions ignite in the degenerate material. Suppose the entire layer
of $\Delta m$ is expelled from the system in a nova explosion. Since
$\Delta M_1=\Delta M =-\Delta m$, the angular momentum of the system
changes by
\begin{equation}
\Delta J=\Delta M_1 a_1^2 \Omega = -\Delta m \left(\frac{aM_2}{M}\right)^2\Omega .
\end{equation}
We want to find an expression for the change in separation during the nova eruption. Equations (8) and (11) give
\begin{equation}
\frac{\Delta J}{J}=-\frac{\Delta m M_2}{MM_1}.
\end{equation}
Differentiating equation~(8) gives
\begin{equation}
\frac{\Delta J}{J}=\frac{\Delta M_1}{M_1}+\frac{\Delta M_2}{M_2}-\frac{\Delta M}{M}+2\frac{\Delta a}{a}+\frac{\Delta \Omega}{\Omega},
\end{equation}
but, because $M_2$ is constant during the eruption, $\Delta M_2=0$.
Differentiating equation~(3) we get,
\begin{equation}
2\frac{\Delta \Omega}{\Omega}=\frac{\Delta M}{M}-3\frac{\Delta a}{a}.
\end{equation}

We can combine equations~(12),~(13) and~(14) to get the familiar result
\begin{equation}
\frac{\Delta a}{a}=\frac{\Delta m}{M}.
\end{equation}
Because of the mass loss from the system, the separation of the stars
increases during the explosion. Equations~(12) and~(15) ensure that
$a(M_1+M_2)=\rm const$ during the eruption. The Roche lobe radius
must also increase as the separation increases. Differentiating
equation~(2) and substituting equation~(15) we find
\begin{equation}
\frac{\Delta R_{\rm L2}}{R_{\rm L2}}=\frac{4\Delta m}{3M}.
\end{equation}
After the eruption $R_2<R_{\rm L2}$ and so the mass-transfer rate decreases
sharply, switches off in this model. In this low mass-transfer rate the cataclysmic binary is
said to be in a state of hibernation.

\subsection{Conservative Case}
We now  consider a fully conservative case for which $M=\rm const$ and $\dot{J}=0$ so
\begin{equation} 
\frac{\dot{J}}{J}=\frac{\dot{M_1}}{M_1}+\frac{\dot{M_2}}{M_2}+\frac{\dot{a}}{2a}=0.
\end{equation}
Since $\dot{M}_1=-\dot{M}_2$ we can substitute to find
\begin{equation}
\frac{\dot{a}}{a}=2(q-1)\frac{\dot{M}_2}{M_2}
\end{equation}
and
\begin{equation}
\frac{\dot{P}}{P}=3(q-1)\frac{\dot{M}_2}{M_2},
\end{equation}
where $q=M_2/M_1$.  These two equations show that, as the mass of the
red dwarf decreases, $a$ and $P$ only decrease if $q>1$. But because
the Roche lobe continues to shrink as $q$ decreases, mass transfer
doesn't end at $q=1$. Differentiating equation~(3) we get
\begin{equation}
\frac{\dot{R}_{\rm L2}}{R_{\rm L2}}=\frac{\dot{a}}{a}+\frac{\dot{M_2}}{3M_2},
\end{equation}
and combining with equation~(18) we find
\begin{equation}
\frac{\dot{R}_{\rm L2}}{R_{\rm L2}}=\frac{\dot{M_2}}{M_2}\left(2q-\frac{5}{3}\right).
\end{equation}
This shows that the minimum Roche lobe radius is when
$q=5/6$. But for most observed cataclysmic binaries $q<5/6$ and steady
mass transfer occurs. For a red dwarf to be in steady contact with its
Roche lobe there must be something causing loss of orbital angular
momentum.

So after an eruption the stars gradually get closer together because
of the loss of angular momentum which brings the main-sequence star
back into contact with its Roche lobe. The mass-transfer rate
increases once more and the cycle is repeated. We have many eruptions.

We want to find the range of mass ratios $q$ for which mass transfer is stable. Since we have, by differentiating equation~(1)
\begin{equation}
\frac{\dot{R}_2}{R_2}=\frac{\dot{M}_2}{M_2}
\end{equation}
and in order to be stable we need
\begin{equation}
\frac{\dot{R}_{\rm L2}}{R_{\rm L2}}>\frac{\dot{R}_2}{R_2},
\end{equation}
by comparing equations~(21),~(22) and~(23) we therefore must have
\begin{equation}
q<4/3.
\end{equation}

\subsection{Analytic Ratio of Times}

After a nova eruption the main-sequence star no longer fills its Roche
lobe because the orbit has expanded.  Here we assume that the rate of
angular momentum loss $|\dot{J}|$ remains constant until the next nova
explosion.

When the system is detached (when there is no mass transfer)
$\dot{M_1}=\dot{M_2}=\dot{M}=0$. So from one nova explosion, the time
until mass transfer begins again is the time it takes until $R_{\rm L2}$
shrinks back to $R_2$. From equations~(13), (14) and~(16),
\begin{equation}
\frac{\Delta J}{J}=\frac{|\dot{J}|t_{\rm d}}{J} =\frac{\Delta a}{2a}=\frac{\Delta R_{L2}}{2R_{L2}}=\frac{2\Delta m}{3M}.
\end{equation}
So the time the system spends detached is,
\begin{equation}
t_{\rm d}=\frac{J}{|\dot{J}|}\frac{2\Delta m}{3M}.
\end{equation}

\begin{figure}
\epsfxsize=8.4cm \epsfbox{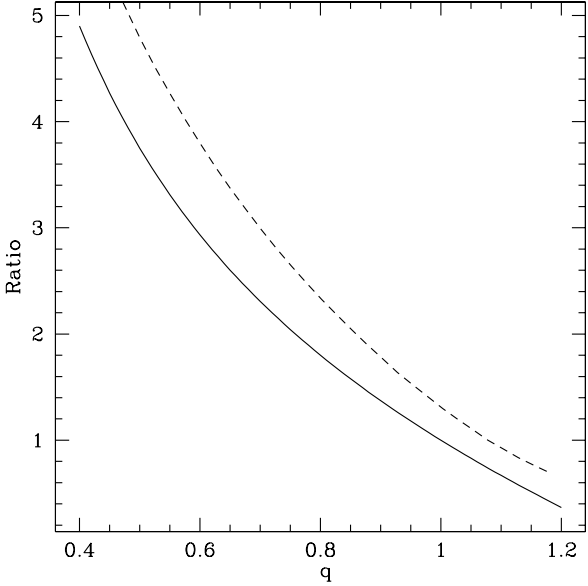}
\caption[]{The solid line is $t_{\rm s}/t_{\rm d}$ against the mass
ratio $q$ (equation~31).  We compare with the dashed line which is the ratio of the time spent as a
nova-like to dwarf nova against $q$ with a constant period $P=3\,$hr,
$\dot J$ factor of $0.1$ and $M_{\rm ign}=M_{\rm ej}$ from our numerical model (section~5).}
\label{tstdq}
\end{figure}

When the system is semi-detached mass is transferred from the red
dwarf to the white dwarf, $\dot{M}_1=-\dot{M}_2$ and $\dot{M}=0$. The
Roche lobe of the red dwarf is filled so that $R_2=R_{\rm L2}$ and
$\dot{R_2}=\dot{R}_{\rm L2}$ so
\begin{equation}
\frac{\Delta{R}_{\rm L2}}{R_{\rm L2}}=\frac{\Delta{a}}{a}+\frac{\Delta{M}_2}{3M_2}=\frac{\Delta{R}_2}{R_2}=\frac{\Delta{M}_2}{M_2}.
\end{equation}
Rearranging this we find
\begin{equation}
\frac{\Delta{a}}{a}=\frac{2\Delta{M}_2}{3M_2}.
\end{equation}
Substituting~(28) into~(13) we get
\begin{equation}
\frac{\Delta J}{J}=\frac{|\dot{J}|t_{\rm s}}{J}=-\frac{\Delta m}{M_1}+\frac{\Delta m}{M_2}+\frac{\Delta m}{3M_2}.
\end{equation}
The time spent semi-detached is
\begin{equation}
t_{\rm s}=\frac{J}{|\dot{J}|}\Delta m \frac{(4-3q)}{3M_2}.
\end{equation}
This model predicts the ratio of time spent detached to the time spent semi-detached to be,
\begin{equation}
\frac{t_{\rm s}}{t_{\rm d}}=\frac{(1+q)(4-3q)}{2q}. 
\end{equation}
We plot it in Fig.~\ref{tstdq} for later comparison with our numerical
model and note that this is consistent with the result of \cite{b5}
that hibernation should be deepest when $q$ is close to one.

\section{Observed Cataclysmic Variables}

\indent \textbf{Classical novae} are a class of novae and cataclysmic
variables that have had only a single observed eruption. Their
brightness ranges from~6 to greater than~19 magnitudes.
\textbf{Dwarf novae} (DNe) are a class of cataclysmic variables that
have had several observed outbursts.  These outbursts however range in
brightness from 2 to~5 magnitudes.  They are associated with a disc
instability.
\textbf{Nova-like systems} (NLs) include all non-outbursting cataclysmic
variables.  Their spectra indicate that they could be novae in a pre- or
post-eruption stage.

The nature of the cataclysmic variable, DN or NL, should depend only
on the mass-transfer rate which itself should only depend on the
system masses and period. However at any given period some of each
type are seen.

\section{Numerical Model}
\indent We now estimate the radius and luminosity of the main-sequence
 star with the formulae of \cite{b2} and use the more accurate formula
 for the Roche lobe radius for star $i$ \citep{Egg83},
\begin{equation}
\frac{R_{{\rm L}i}}{a}=\frac{0.49q_i^{2/3}}{0.6q_i^{2/3}+\log(1+q_i^{1/3})} ,
\end{equation}
where $q_i$ is the mass ratio $M_i/M_{3-i}$. This formula is valid for all $q$.

\subsection{Ignition Mass}

By considering cold white dwarf models \cite{Fuj} calculated the
critical pressure at the base of the hydrogen rich layer for
ignition. Based on this \cite{Tru} derived
\begin{equation}
\frac{M'_{\rm ign}}{M_\odot}=\frac{9.42\times10^{-4}\left( R_1/10^9
R_\odot \right)^4}{M_1/M_\odot},
\end{equation}
where the prime distinguishes this from a prescription that includes
the effect of varying the mass-transfer rate.

However the white dwarf is heated by accretion so the rate also
depends on $\langle \dot M \rangle$.  \cite{TandB} have calculated the
critical mass which causes ignition on the white dwarf. We make a
linear fit to their figure~8 and find
\begin{equation}
\frac{M_{\rm ign}}{M_\odot}=\alpha \left(\frac{\langle\dot M\rangle}{10^{-8}\,{ M_\odot \,\rm yr^{-1}}}\right)^{-\beta},
\end{equation}
where
\begin{equation}
\alpha=3.2\times 10^{-4}\left(\frac{M_\odot}{M_1}\right)^{1.231}
\end{equation}
and
\begin{equation}
\beta = 0.534 \left(\frac{M_1}{M_\odot}\right)^{0.605}.
\end{equation}
When $M_{\rm ign}$ has accumulated on the white dwarf we have a nova
explosion. We consider three different models, two in which the mass ejected is
constant in all the explosions,
\begin{enumerate}
\item $M_{\rm ej}=\rmn{const}=1\times10^{-3} M_\odot$ 
\item $M_{\rm ej}=\rmn{const}=5\times10^{-4}M_\odot$ 
\end{enumerate}
and one in which the mass ejected in the nova equals the amount of
mass needed for ignition,
\begin{enumerate}
\item[(iii)] $M_{\rm ej}=M_{\rm ign}$.
\end{enumerate}

\subsection{Rate of loss of Mass}
\indent 

When the main-sequence star is overfilling its Roche lobe mass is transferred on to the white dwarf so that
\begin{equation}
\dot{M_1}=-\dot{M_2}.
\end{equation}
We approximate the atmosphere of the red dwarf by a simple isothermal
model. The rate of mass loss from the red dwarf is then given by,
\begin{equation}
\dot{M_2}=-M_2\left(\frac{GM_2}{R_2^3}\right)^{1/2} e^{\frac{\Delta R}{H}},
\end{equation}
where $\Delta R$ is the amount by which the star overfills its Roche lobe, 
\begin{equation}
\Delta R= R_2-R_{\rm L2}.
\end{equation}
When the red dwarf underfills its Roche lobe, $R_2<R_{\rm L2}$, $\Delta
R$ is negative and the mass-transfer rate is very small. When the star
overfills its Roche lobe, $R_2>R_{\rm L2}$, $\Delta R>0$ and the mass
transfer rate increases.
 
The scale height $H$ of the unperturbed isothermal atmosphere of the main-sequence star is
\begin{equation}
H=-\frac{\rho(r)}{\frac{d\rho(r)}{dr}},
\end{equation}
so we need to find $\rho(r)$, the density at position $r$.
The equations of an isothermal atmosphere are
\begin{equation}
\frac{dP(r)}{dr}=-\rho(r) g \qquad \mbox{and} \qquad
P(r)=\frac{{\Re}T_{\rm e}}{\mu}\rho(r),  \label{des}
\end{equation}
where $P(r)$ is the pressure, $\Re$ is the gas constant, $T_{\rm e}$ is
the effective temperature at the surface of the main-sequence star and
$\mu$ is the mean molecular weight. The atmosphere is a mixture of
hydrogen, helium and metals in various states of ionization and in
molecules. So we use $\mu$ equal to $1$. The gravitational field
strength at the surface of the red dwarf is,
\begin{equation}
g=\frac{GM_2}{R_2^2}.
\end{equation}
The surface temperature of the main-sequence star is found by Stefan's law,
\begin{equation}
T_{\rm e}= \left(\frac{L_2}{4\pi \sigma R_2^2}\right)^{1/4}.
\end{equation}
Integrating the equations in~(\ref{des}) for $\rho$,
\begin{equation}
\rho=\rho_o e^{-\frac{\mu g}{\Re T_{\rm e}}r},
\end{equation}
where $\rho_o$ is some constant.
Hence we find the scale height,
\begin{equation}
H=\frac{\Re T_e}{\mu g}.
\end{equation}
The factor in front of the exponential in equation~(38) is an estimated
maximum mass-transfer rate equal to the mass of the secondary star
divided by its dynamical timescale. We have verified that varying it
by large factors has almost no effect on our results because $\Delta
R$ simply adjusts to compensate when mass transfer is stable. 

\subsection{Rate of Loss of Angular Momentum}

Angular momentum is continually lost from the system. For the closest
systems gravitational radiation is the most important cause of mass
transfer. The rate of loss of angular momentum in gravitational
radiation from two point masses in a circular orbit is \citep{landau1951}
\begin{equation}
\frac{\dot{J}_{\rmn{GR}}}{J}=-\frac{32G^3}{5c^5}\frac{M_1M_2(M_1+M_2)}{a^4}.
\end{equation}

In wider systems however magnetic braking is the more important
mechanism.  The rate of change of angular momentum owing to magnetic
braking is uncertain.  Based on the work of \cite{skum}, \cite{Rapp}
postulated a rate of the form

\begin{equation}
\label{MB}
\dot{J}_{\rmn{MB}}=-5.83\times10^{-16} (R_2/R_\odot)^3 (\Omega\,{\rm
yr})^3\,{M_{\odot}\,R_{\odot}^2\,\rm yr^{-2}}.
\end{equation}

However recent work by \cite{And} based on the spins of stars in
clusters suggests that this rate is orders of magnitude too high. But
such a low rate would not be able to drive the interrupted magnetic
braking model for the period gap between two and three hours in
cataclysmic variables. In this model a system evolves to lower period
as $M_2$ drops at such a high rate that the red dwarf is out of
thermal equilibrium. At a period of three hours, a mass of about $0.3
M_\odot$, magnetic braking abruptly falls off, perhaps because the red
dwarf has become fully convective. The mass-transfer rate drops and
the red dwarf shrinks back to its equilibrium radius and must wait for
some much weaker angular momentum loss mechanism to shrink the orbit
until $R_{\rm L2}$ is again reduced to $R_2$. With the high rate the
period gap is very well reproduced. With the low rate the stars are
not far enough from thermal equilibrium and, even if magnetic braking
were interrupted, the gap would be much too narrow. If such a low rate
proves to be correct an alternative reason for the period gap must be
found.

\subsection{Irradiation}
The nova decay time is $50$ to $100\,\rm yr$ \citep{b5}. So although
irradiation of the red dwarf from the white dwarf may increase the
mass-transfer rate initially, it will only be for a small fraction of
the inter nova period of several thousand years. Though we find
increased mass transfer while the white dwarf is hot the time for
which it is is negligible and we ignore it. We note that the
irradiation cycles discussed by \cite{bun} operate on a much longer
timescale and would have to survive several episodes of hibernation.

\section{Ratio of Number of Observed NLs to DNe}
We select the systems from Ritter's Catalogue \citep{Ritt} which are
classified as nova-likes or as dwarf novae and divide them into period
bins with at least five systems of each kind in each bin. We then
divide these numbers to find the observed ratio of NLs to DNe. Though
there are obvious selection effects, such as the fact that dwarf
novae, owing to their variability, are far more easily spotted, we can
hope to reproduce trends in this ratio.

\begin{figure}
\epsfxsize=8.4cm \epsfbox{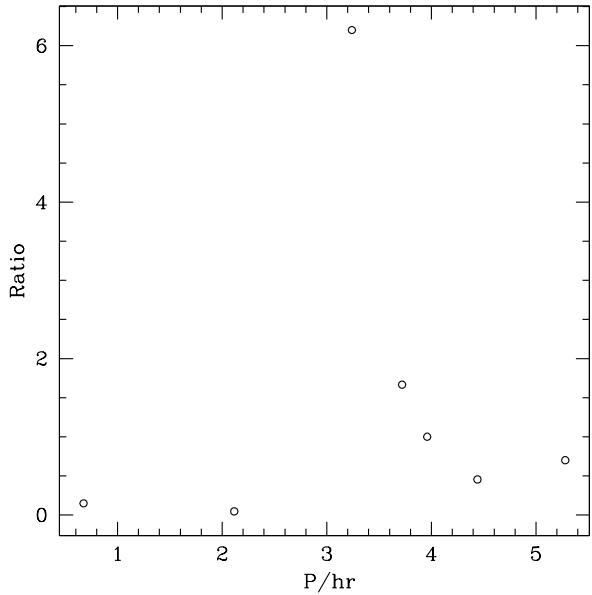}
\caption[]{The ratio of the number of observed nova-likes to dwarf
novae against the period.  Each point is a ratio based on at least
five of each type of system.  Poisson noise error bars are therefore
typically a factor of two of the height of the point.}
\label{obs}
\end{figure}

Fig.~\ref{obs} shows that the ratio peaks at a period of just over
$3\,$hr then decays as the period increases further. After a period of
$5\,\rm hr$ there are very few observed binaries but the data do
suggest that the ratio increases again slightly with increasing
period.  We use the period of the peak ratio $3.5\,\rm hr$ and the fact
that it has dropped by a factor of twelve by $4.4\,\rm hr$ to test our
models.  We do not expect to reproduce the exact height of the peak
because different selection effects apply to DNe and NLs.  We
exclude identified polars and intermediate polars but we cannot be certain
that others, particularly IPs, do not contaminate the sample of NLs.
However their contribution is small, certainly when compared with
other probable selection effects, at the relatively high periods we
consider here.

\subsection{Critical Mass-Transfer Rate}
\begin{figure}
\epsfxsize=8.4cm
\epsfbox{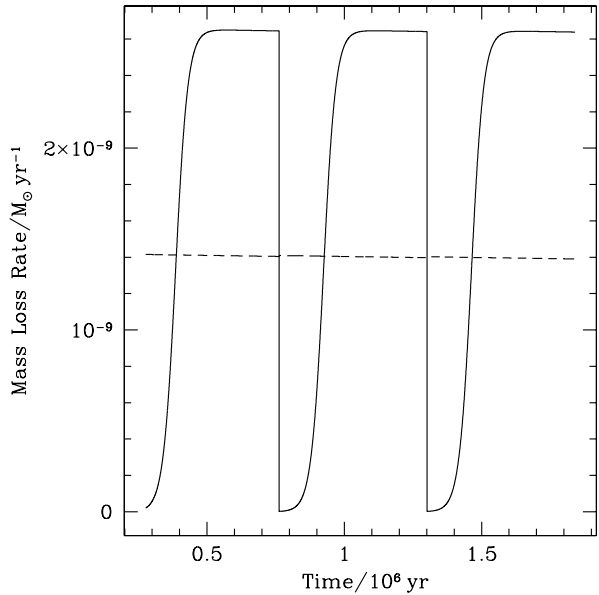}
\caption[]{The mass-loss rate against time for a system with initial
mass ratio $q=0.8$ and period $P=3.5\,\rm hr$ with $M_{\rm ej} =
M_{\rm ign}$ and $\dot J = 0.1\dot J_{\rm MB}$ in
equation~(\ref{MB}). The dashed line is the critical rate below which
the system would undergo dwarf nova outbursts and above which it would
not.}
\label{mdot}
\end{figure}

We calculate the times spent as a dwarf nova $t_{\rmn{DN}}$ and as a
nova-like $t_{\rmn{NL}}$ by considering the mass-transfer
rate. Fig.~\ref{mdot} shows an example of the mass-loss rate of the
red dwarf for three  nova explosions.

There is a critical mass-transfer rate above and below which the
system appears to be in two different states. This critical rate of
mass loss from the main-sequence star is
\begin{equation}
\dot{M_2}_{\rmn {crit}}=-8.08\times10^{15}\left(\frac{\alpha
_H}{0.3}\right)^{0.3}(1+q)^{7/8}\left(\frac{P}{\rm
hr}\right)^{7/4}\rmn{\,g\, s^{-1}}  \label{mlr}
\end{equation} \citep{Faulk},
where $\alpha _H=0.1$.  Dwarf novae lie below the critical
value. They have $\langle\dot{M_2}\rangle <\dot{M}_{2\rmn{crit}}$
whereas nova-likes all lie above the critical $\dot{M}_{2\rmn{crit}}$
relationship.

Just after a nova explosion, the mass-loss rate from the red dwarf is
very small. The time from the explosion until the mass-loss rate has
increased up to the critical rate is $t_{\rmn{DN}}$. The binary then
appears to be a nova-like from this time until the next eruption,
$t_{\rmn{NL}}$. We calculate the ratio of these times.
\begin{equation}
\rmn{Ratio}=\frac{t_{\rmn{NL}}}{t_{\rmn{DN}}}.
\end{equation}

 If the critical mass-transfer rate is higher than the rate of the
model for the whole cycle, then we are predicting that there are no
nova-likes, only dwarf novae.  So we have $t_{\rmn{NL}}=0$ $
t_{\rmn{DN}}=$ time between nova eruptions. Similarly, if the critical
rate is lower than the mass-transfer rate, we have only nova-likes,
$t_{\rmn{DN}}=0$ and $ t_{\rmn{NL}}=$ time between nova eruptions.
\par
In Fig.~\ref{tstdq} we compare this ratio for $P = 3\,$hr against $q$
with the simple analytic calculation in section 2.4. At a period of
$3\,$hr there are both NLs and DNe at approximately all mass ratios
and we see that both models are indeed very similar.

\subsection{Model Ratio Averaging}
In order to compare our models with the observed data we need to plot
the predicted ratio against the period but the ratio also depends on
$q$. For a given period, we average the ratio over varying
$q$. Averaging by adding the total times, $t_{\rm NL}$ and $t_{\rm
DN}$, over a flat $q$ distribution and then finding the ratio of these total
times is our preferred method. It automatically takes account of system
lifetimes. The longer the time, the more weight it is given in the
averaging. In the observed data, the longer a system is in a state,
the more likely it is to be observed. We do not try to model the birth
rate of various systems with population synthesis because we consider
this to be too uncertain. We therefore implicitly assume all
cataclysmic variables equally likely to form.

\section{Results}
To examine the effects of a much lower angular momentum loss rate we
multiply the formula for magnetic braking (equation 47) by a
constant factor. We try combinations of different
$\dot{J}$ factors and $M_{\rm ej}$ and compare the ratios in Table~$1$.

\begin{table}
 \caption{Models with different factor in $\dot J$ and $M_{\rm ej}$}
 \label{symbols}
 \begin{tabular}{@{}lcccccc}
  \hline
  $\dot{J}$ Factor & $M_{\rm ej}/M_\odot$ & Peak & Peak & Ratio after peak \\
  && Period/hr & Ratio &  at $P=4.4\,\rm hr$\\

  \hline
  1.0 & $M_{\rm ign}$    & 4.5-7.25 & inf & -\\            
  1.0 & $5\times 10^{-4}$ & 6.0 & inf  & -\\
  1.0 & $1\times10^{-3}$ & 3.5 & 4.7 & 4.3 \\
  0.5 &  $M_{\rm ign}$ & 5.75 & 13.5 & - \\
  0.2 &  $M_{\rm ign}$ & 3.25 & 3.1 & 2.0\\
  0.15 & $M_{\rm ign}$ & 3.25 & 2.8 & 1.0 \\
  0.1 & $M_{\rm ign}$ & 3.0 & 2.4 & 0.2 \\
  0.1 & $5\times 10^{-4}$ & $<$2.0 & $>$4.0 & 0.2\\
  0.1 & $1\times10^{-3}$ & $<$2.0 & $>$3.4 & 0.2\\
  0.1 & $M'_{\rm ign}$ & $<$3 & $>$1.2 & 0.6\\
 
  \hline
 \end{tabular}
 \medskip
\end{table}

In Table~1 we list various models and properties which can be compared
with the observations. As we have discussed, selection effects make
any truly quantitative comparison difficult but we have identified
key features in Fig.~\ref{obs} that ought to be reproduced. We also
concentrate on periods above the gap because we are interested in what
can be said about magnetic braking irrespective of whether or not it
is interrupted. Then we see a peak in the ratio of NLs to DNe just above
$3\,\rm hr$ with a rapid decrease to higher periods.  By $4.4\,\rm
hr$ it should have fallen by a factor of~12.  A record of `inf' for the
peak ratio indicates no dwarf novae.

With the full \cite{Rapp} braking rate ($\dot J$ factor = 1) we cannot
reproduce these features. When just the accreted mass is blown off
there are no DNe at all between about $4.5$ and~$7.25\,\rm hr$. If a
constant $5\times 10^{-4}M_\odot$ mass is ejected then there is a peak
in the ratio at $6\,\rm hr$. Only when $M_{\rm ej}=10^{-3}M_\odot$ do we find
a peak at about $3.5\, \rm hr$ but then the ratio has hardly fallen by
$4.4\,\rm hr$. In any case such a large ejected mass is unrealistic.
\cite{kolb2001} investigated the effects of hibernation at these
standard rates of magnetic braking.  They deduced that $M_{\rm ej} >
2\times 10^{-4}\,M_\odot$ would be necessary to account for the observed
spread in $\dot M$ at a given period above $3\,$hr.

It is not until we have reduced the braking rate by a factor of
ten that we can reproduce the observed features.  There is little
variation between the three different $M_{\rm ej}$, all of which show
a peak in the ratio at about $3\,\rm hr$ with a drop by a factor of
five or more by $4.4\,\rm hr$. We also made one model with the old
cold white dwarf ignition mass $M_{\rm ej}=M'_{\rm ign}$. Here, the
ratio is only $1.2$ at $P=3\,\rm hr$ predicting about equal numbers
of NLs and DNe. Because DNe should be easier to observe than NLs this
is inconsistent with the data too.

\begin{figure}
\epsfxsize=8.4cm
\epsfbox{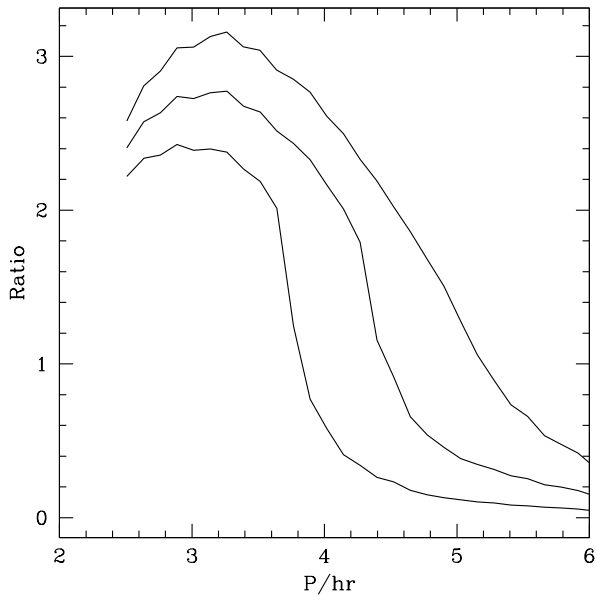}
\caption[]{The ratio against period for the models with
$\dot{J}\rm\,factor =0.1, 0.15, 0.2$ (from bottom to top) and $M_{\rm
ej}=M_{\rm ign}$.}
\label{ratP}
\end{figure}

Fig.~\ref{ratP} shows the ratios as a function of period for $\dot J$ factors
of 0.1, 0.15 and 0.2. All have the correct shape but as the $\dot J$
factor increases the curves flatten, no longer reproducing the drop of
a factor of $12$ by $4.4\,\rm hr$.

For a $\dot{J}$ factor of 0.01, which gives a rate close to that of
\cite{And}, all of the mass-transfer rates are lower than the critical
value and so we see only dwarf novae. The only way this could be
consistent with observations would be if another mechanism such as
irradiation or magnetic cycles can periodically raise the mass
transfer by an order of magnitude. This would then be responsible for the
variety of cataclysmic variable types.

\section{Conclusions}

For the full \cite{Rapp} rate ($\dot J$ factor of $1$) only the
unrealistically high $M_{\rm ej}$ of $10^{-3}M_\odot$ follows the
observed trend but then the variation of ratio with period is too small.

With $\dot J$ reduced by a factor of ten all the models show the right
trends with the ratio rising to a peak at a period of about $3\,\rm
hr$ and then falling distinctly by a period of $4.4\,\rm hr$. We
cannot then distinguish between the different mass ejection models but
do find a distinct improvement when $M_{\rm ej}=M_{\rm ign}$ when
proper account is taken of white dwarf heating by accretion
\citep{TandB}. Hibernation is most effective at reproducing observed
trends when our $\dot J$ factor is about $0.1$. This is ten times
smaller than that proposed by \cite{Rapp} and ten times larger than
that proposed by \cite{And}.

\section*{Acknowledgements}
We thank Friedrich Meyer and Phillip Podsiadlowski for useful
conversations. We are grateful to Melvyn Davies and Lund Observatory
for their hospitality during which a substantial amount of this work
was carried out. RGM thanks Steven Smith for his assistance in
interrogating Ritter's Catalogue. CAT thanks Churchill College for a
fellowship.

\label{lastpage}
\end{document}